\documentclass[utf8]{FrontiersinHarvard}

\usepackage{url,hyperref,lineno,microtype,subcaption}
\usepackage[onehalfspacing]{setspace}
\usepackage{gensymb}
\usepackage{xcolor}
\usepackage{ulem}

\def\keyFont{\fontsize{8}{11}\helveticabold }
\def\firstAuthorLast{Mazur {et~al.}}
\def\Authors{Michael J. Mazur\,$^{1,*}$, Stanimir Metchev\,$^{1,2}$ Rachel A. Brown\,$^{1,2}$, Ridhee Gupta\,$^{3}$, Richard Bloch\,$^{4}$, Tristan Mills\,$^{1}$, Emily Pass\,$^{5}$}

\def\Address{$^{1}$Western University, Department of Physics and Astronomy, London, Ontario, Canada\\

{$^{2}$Western University, Institute for Earth and Space Exploration, London, Ontario, Canada\\}

{$^{3}$University of Waterloo, Department of Physics, Waterloo, Ontario, Canada\\}

$^{4}$York University, Department of Physics and Astronomy, 4700 Keele Street, Toronto, Ontario, Canada\\

$^{5}$Center for Astrophysics $\vert$ Harvard \& Smithsonian, 60 Garden Street, Cambridge, Massachusetts, USA}

\def\corrAuthor{Dept. of Physics and Astronomy, Western University, 1151 Richmond St., London, Ontario, Canada N6A 3K7}

\def\corrEmail{mmazur5@uwo.ca}

\begin{document}
\onecolumn
\firstpage{1}

\title[The Colibri KBO Observatory]{The Colibri Telescope Array for KBO Detection through Serendipitous Stellar Occultations: a Technical Description} 

\author[\firstAuthorLast ]{\Authors}
\address{}
\correspondance{}

\extraAuth{}

\maketitle

\begin{abstract}

\section{}
We present the technical design, construction and testing of the Colibri telescope array at Elginfield Observatory near London, Ontario, Canada. Three 50-cm telescopes are arranged in a triangular array and are separated by 110-160 metres. During operation, they will monitor field stars at the intersections of the ecliptic and galactic plane for serendipitous stellar occultations (SSOs) by trans-Neptunian objects (TNOs). At a frame rate of 40 frames per second (fps), Fresnel diffraction in the occultation light curve can be resolved and, with coincident detections, be used to estimate basic properties of the occulting object. Using off-the-shelf components, the Colibri system streams imagery to disk at a rate of 1.5 GB/s for next-day processing by a custom occultation detection pipeline.

The imaging system has been tested and is found to perform well, given the moderate site conditions. Limiting magnitudes at 40 fps are found to be about 12.1 (temporal SNR=5, visible light Gaia $G$ band) with time-series standard deviations ranging from about 0.035 mag to \textgreater0.2 mag. SNR is observed to decrease linearly with magnitude for stars fainter than about $G = 9.5$ mag. Brighter than this limit, SNR is constant, suggesting that atmospheric scintillation is the dominant noise source. Astrometric solutions show errors typically less than $\pm$0.3 pixels (0.8 arc seconds) without a need for high-order corrections.

\tiny
 \keyFont{ \section{Keywords:} robotic telescopes, Kuiper Belt Objects, occultation, photometry, Fresnel diffraction}
\end{abstract}

\section{Introduction}
\subsection{The Undiscovered Population of Kilometre-Sized Outer Solar System Objects}
Beyond Neptune orbits a population of up to $10^{11}$ objects with sizes of a kilometre or larger \citep{Roques2000}, known as trans-Neptunian objects (TNOs). It can be divided into three main subgroups: Kuiper Belt objects (KBOs), scattered disk objects, and the detached objects such as the Sednoids and the distant Oort Cloud objects \citep{Delsanti2006}. Extending from about 30 AU to about 50 AU from the sun, the Kuiper Belt contains objects with predominantly low inclinations that are dynamically stable compared to the broader group of TNOs. The KBOs can be subdivided into two groups known as the dynamically cold and hot populations. The cold population typically have low inclination (\textless5\textdegree) orbits and are believed to represent a dynamically primordial population \citep{Levison2001}. The cold population is also expected to have the highest sky surface density, since it most closely follows the solar system ecliptic plane.

Compared to asteroids, for which the size-frequency distribution is reasonably well understood, less is known about this population of distant solar system objects. Because of their distance from Earth, smaller KBOs can be difficult or impossible to observe directly. Of the more than one thousand KBOs discovered by direct imaging, only a small fraction (3\%) are less than 25 km in size and none are less than 7 km. 
As a result, the size-frequency distribution of KBOs and, more generally TNOs, is poorly defined for smaller objects.

\subsection{Indirect Detection through Serendipitous Stellar Occultations}

Using stellar occultations as a method for the indirect detection of `invisible' bodies in the Solar System was suggested by \citet{Bailey1976}. Using a 1 m telescope with a vidicon imager and a visual ($V$-band: $\approx$505--595~nm) limiting magnitude of 16, Bailey suggested that a 1000-star field would have an occultation event every 11 hours. His analysis, however, is optimistic as it does not consider the effects of line-of-sight Fresnel diffraction nor fully explore the temporal limit imposed by a 10 Hz sample rate.

A theoretical description of Fresnel diffraction during stellar occultations by small bodies was discussed by \citet{Roques1987}. 
Fresnel diffraction is used to describe near-field diffraction and, given the correct geometries and source/object sizes, can be used to describe the observed light curves from KBO occultation events. \citet{Roques1987} examined both the theory and implementation of diffraction models and forms the basis of subsequent observational studies in the field. 
\citet{Roques2000} further explored the possibility of detecting small bodies in the Solar System. Their work focused on sub-kilometre KBOs and predicted the possibility of a few to several tens of occultation events per night with 2-m to 8-m class telescopes at a visual limiting magnitude of 10 mag $\leq V \leq$ 12 mag. With today's instrumentation, however, similar limiting magnitudes are achievable with smaller telescopes operating at higher frame rates.

In order to sufficiently resolve a KBO occultation light curve, we need to image at relatively high frame rates. For a 1 km KBO at 40 AU observed at solar opposition, its speed relative to the observer is about 25 km/s. If this object occults a star with a 1 km projected diameter, the occultation will have a duration $\Delta T \approx 80$ ms if we consider the event purely geometrically. However, the diffractive broadening of the geometric shadow prolongs the event by a factor of 2--3. The effect is more pronounced for smaller stellar disks and better photometric precision \citep{Roques1987, Roques2000}.

Taking into account relative velocities, Fresnel geometry, and telescope sensitivity, \citet{Bickerton2009} find that the optimal sampling rate for detecting serendipitous stellar occultations by KBOs at visible wavelengths is 40 fps, with observations toward solar opposition. These are the parameters that we use to set the technical requirements for the Colibri telescope array. These parameters are preliminary. Until reliable KBO occultation detections become routine, the optimal trade-off between telescope size, sampling rate, observing wavelength, and observing geometry has yet to be empirically validated.

Indeed, the \citet{Bickerton2009} study details a much longer set of assumptions to project KBO occultation detection rates. We defer a discussion of the expected event rate specific to the Colibri experiment to a future publication (Metchev et al.\ 2022, in preparation). In the meantime, we note that some recent and on-going experiments (Section~\ref{sec:similar_experiments}) validate our choice of telescope system and operating mode.

\subsection{Other Previous or Planned Experiments}
\label{sec:similar_experiments}

Several prior experiments have reported serendipitous stellar occultations by KBOs. These have often been based on sub-optimal data sets, in some cases acquired for a different science goal. \citet{Chang2006} analyse 90 hours of archival x-ray monitoring observations of a single source, Cygnus X-1, with the Rossi X-ray Timing Explorer (RXTE) satellite at 500 fps. The 58 candidate occultations reported in their data are by far the largest in a single data set. However, subsequent re-analyses of the data and their statistical interpretation by \citet{Jones2006} and \citet{Bickerton2008} have put these detections into doubt. \citet{Schlichting2009, Schlichting2012} report two different candidate occultations from visible-light guiding operations at 40 fps from over 20 years of observatinos with the Hubble Space Telescope. These events do bear the hallmarks of the expected Fresnel diffraction pattern of stellar occultations by kilometre-sized KBOs \citep[e.g., Figure 1 of][]{Schlichting2009}. Detections of similar events with other facilities would confirm them as representative of this phenomenon.

Early observations designed specifically for the detection of serendipitous stellar occultations have also yielded some candidate detections and mixed results. \citet{Roques2006} report three candidate events in 11 h of dual-band visible wavelength monitoring of two stars at 45 fps with the 4.2 m William Herschel Telescope. \citet{Bickerton2008} discuss that while the rate of events in this study is in line with expectations, their statistical significance is low. More recently, \citet{Arimatsu2019} report a single candidate event using a pair of 28~cm amateur optical telescopes, in a 60 h observation at 15.4 fps in the course of the Organized Autotelescopes for Serendipitous Event Survey \citep[OASES;][]{Arimatsu2017}. The sub-optimal (\textless40 fps) cadence of the observations and the low (\textless10) SNR of the four individual measurements that constitute the candidate event leave enough room for it to be a false positive. Nonetheless, the OASES setup and its use of commercial hardware and a rapid-imaging complementary metal–oxide–semiconductor (CMOS) camera are promising for designing large-scale serendipitous stellar occultation surveys.

Most significantly, the Taiwanese-American Occultation Survey \citep[TAOS;][]{Lehner2009} was specifically designed to identify $\sim$1 km-diameter objects beyond the orbit of Neptune, and to measure the size distribution of KBOs with diameters between 0.5--30 km. Seven years of visible-light monitoring with initially three, and then four 50 cm telescopes with TAOS yielded no occultation detections \citep{Zhang2013}. This was attributed to a lower-than-expected event rate, and also the relatively slow (5 fps) sampling of the cameras. A follow-up experiment, TAOS II \citep{Lehner2012}, is in the final stages of development, and will use three 1.3 m telescopes imaging at 20 fps. Much like OASES, and the herein described Colibri Telescope Array, TAOS II will use a CMOS-type visible-light camera.

\section{Colibri Hardware}
\begin{figure}
    \centering
    \includegraphics[width=\textwidth]{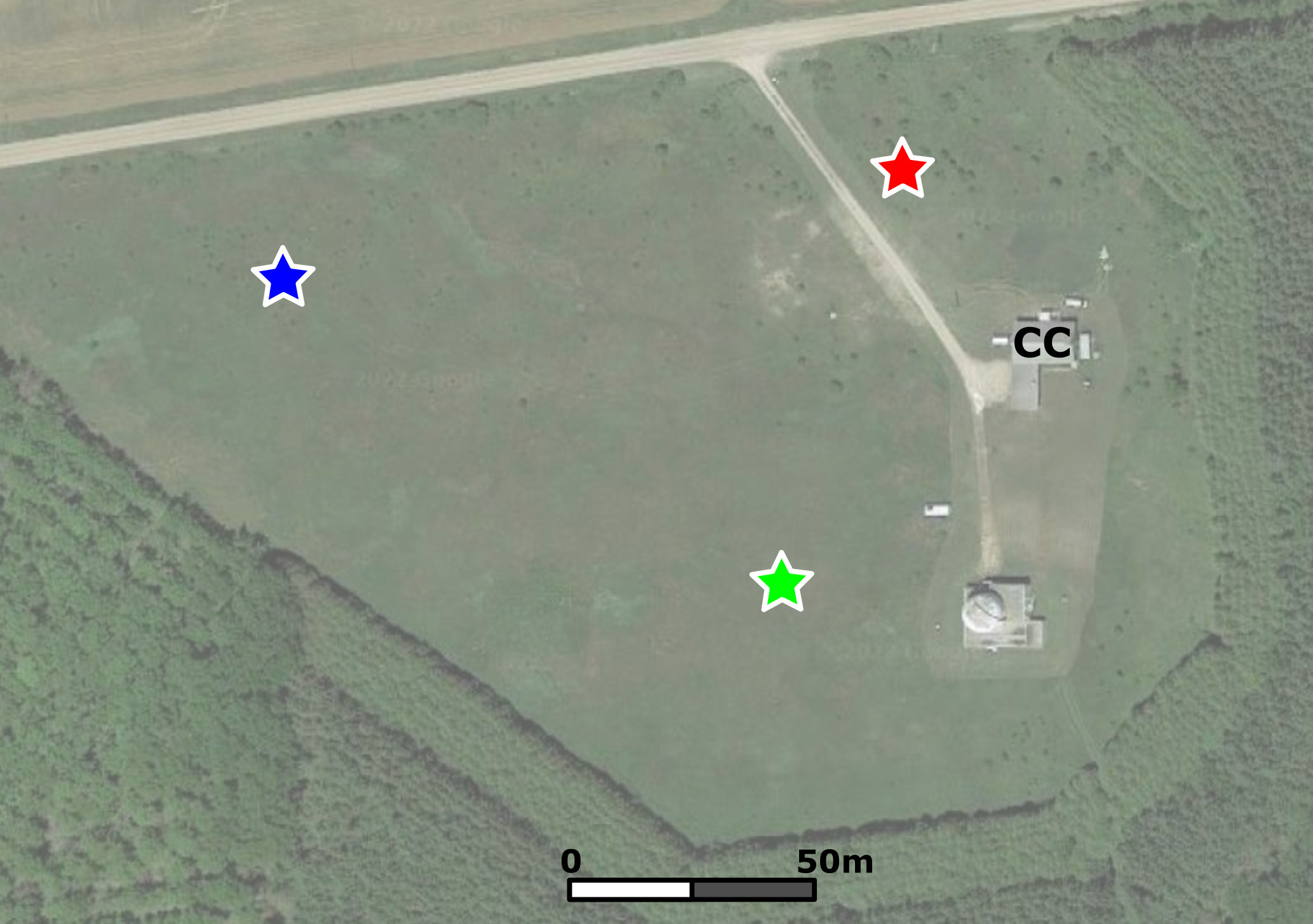}
    \caption{A satellite image showing the location of the three Colibri telescopes (stars) and the control center (CC) located in the basement of a house on the site. Maps Data: Google. Imagery \textcopyright 2022 First Base Solutions, Maxar Technologies.}
    \label{fig:ColibriLayout}
\end{figure}

Since stellar occultations by KBOs tend to be short-lived, they are difficult to observe. Any experiment to identify these transient events needs to not only rapidly (at 40 fps) image background stars but should also provide a mechanism for confirmation. The three telescopes of the Colibri array are set up in a triangular pattern with spacings of between 110 m and 160 m from each other (Figure \ref{fig:ColibriLayout}). This arrangement allows us to rule out the possibility of atmospheric twinkling for coincident events but is not sufficient for size/distance determination by differential transit timing. A full description of the technical specifications of the Colibri array is given in Table \ref{tab:ColibriSpecs}.

\begin{table}
    \caption{Description of the three-telescope Colibri array. }
    \centering
    \begin{tabular}{l c}
    \hline
    & Specification\\
    \hline
    Location 1 (Lat., Long.) & 43.193365\textdegree N, 81.316090\textdegree W\\
    Location 2 (Lat., Long.) & 43.192415\textdegree N, 81.316467\textdegree W\\
    Location 3 (Lat., Long.) & 43.193108\textdegree N, 81.318033\textdegree W\\
    Telescope baselines & 110--160 m \\
    Telescopes & Hercules 50 cm f/3 w/ Wynne Correctors \\
    Cameras & FLI Kepler KL4040\\
    Sensors & GSense 4040\\
    Pixels & 4096x4096, 9$\mu$m pitch\\
    Field-of-View & 1.43\textdegree x1.43\textdegree \\
    Pixel Scale ($2\times2$ binning) & 2.52$''$  per binned pixel \\
    Digitization & low/high gain @ 12-bit\\
    Readout Noise & 3.7 $\mathrm{e^-}$\\
    Dark Current @ -20\textdegree C & \textless0.15 $\mathrm{e^-}$/pixel/s\\
    Typical Gain & low= 19 $\mathrm{e^-}$/ADU, high= 0.8 $\mathrm{e^-}$ /ADU \\
    Quantum Efficiency & \textgreater70$\%$  (460 nm -- 680 nm) \\
    Framerate @ $2\times2$ binning & \textgreater40 fps \\
    Mag. Limit (SNR=3) in 25 ms & $G=12.5$ \\
    Mag. Limit (SNR=5) in 25 ms & $G=12.1$ \\
    \hline
    \end{tabular}
    \label{tab:ColibriSpecs}
\end{table}

\subsection{Telescopes and Domes}
The light-weight carbon fiber optical tube assemblies support 50 cm f/3 cellular mirrors cast from Schott Boro 33 glass. The telescope manufacturer is Hercules Telescopes, which both built the OTAs and cast the mirrors. The mirrors were then ground and figured by an external contractor. ASA Wynne correctors correct field aberrations for imaging at prime focus. FLI Kepler KL4040 cameras are attached to a focus assembly that is controlled by a Seletek Platypus controller. The controller is connected to our local VLAN where it is accessible to the control computers. Absolute timing is achieved by a Garmin GPS attached to each camera.

\begin{figure}
    \centering
    \includegraphics[width=\textwidth]{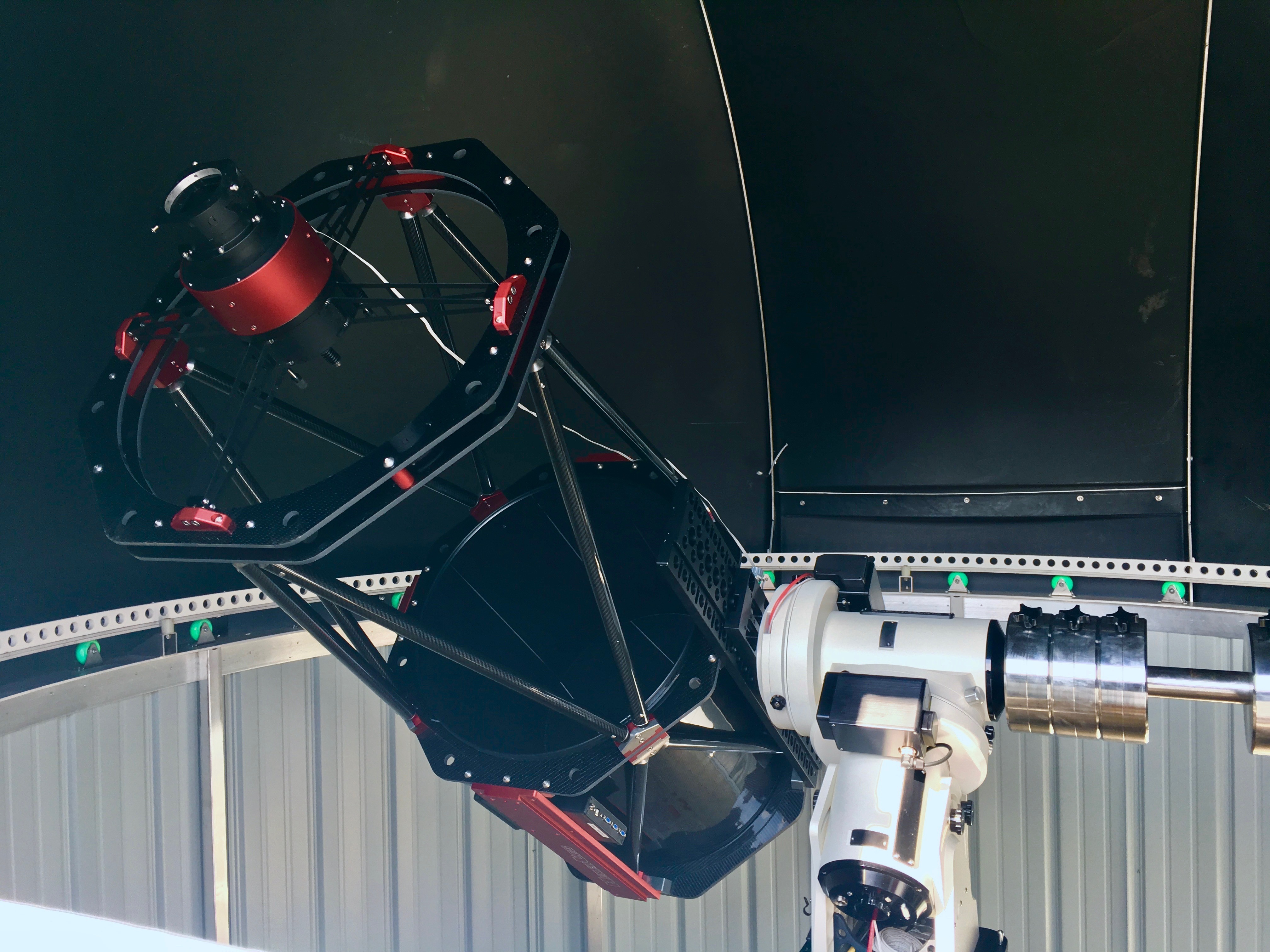}
    \caption{The Hercules 50-cm telescope is mounted on an AP1600 GTO mount inside of a 12$^\prime$ automated dome.}
    \label{fig:ColibriTelescopePhoto}
\end{figure}

The telescopes are attached to Astro-Physics AP1600 GTO equatorial mounts that sit on custom-built steel piers (Figure \ref{fig:ColibriTelescopePhoto}). The steel piers are, in turn, attached to concrete piers extending to about 6 metres below the dome floor. Protection from the elements is provided by ExploraDome EDII 12$^\prime$ plastic domes that have a number of modifications to increase their reliability and allow them to be operated robotically, under the icy and windy conditions common at the site. The original dome control electronics proved to be unreliable and so were replaced with MaxDome II controllers. The dome controllers are connected to USB-over-IP appliances for dome control over much greater distances than would otherwise be possible.

Environmental conditions are monitored on-site by a number of different systems. A central Davis Pro weather station provides an overview of the weather while each dome will be equipped with its own networked environmental station that measures inside/outside temperatures and humidities as well as ambient light levels. Additionally an Arduino-based IR cloud monitor and a dedicated camera continuously aimed at the star Polaris give feedback on seeing conditions. Python scripts talk to each of the devices, parsing and collating the data into a format that can be read by the telescope control software.

\subsection{Imaging Cameras}
Traditional charge-coupled device (CCD) cameras -- available in large formats at reasonable costs -- are still a common option for many astronomical imaging applications. Although most off-the-shelf solutions use high-speed USB interfaces, their frame rates are typically no faster than a few frames per second. As this is still too low for resolving most KBO occultations, commercial CCD cameras do not meet our requirements.

There are other cameras, using electron-multiplying CCDs (EMCCDs), that often have higher frame rates and better noise characteristics than regular CCD cameras. By using an electron-multiplying stage prior to their output amplifier, EMCCD cameras can achieve gains of more than 1000 while maintaining a read noise on the order of a few electrons. Their cost, however, tends to be high and their sensor sizes small. A high-sensitivity EMCCD camera borrowed from the Western Meteor Physics Group was tested, but its high cost, moderate frame rate (17 fps), and small chip dimensions (13 mm x 13 mm) effectively ruled it out as a viable option. The testing did, however, allow the Colibri processing pipeline to be validated on real data and reinforce the need for higher frame rates and a large field of view \citep{Pass2018}.

Over the past few years, there has been a shift within the imaging community from CCD sensors to CMOS sensors. While CMOS-based cameras have been found in cameras and mobile phones for years, scientific CMOS (sCMOS) cameras are relatively new to the astronomical market. Their architecture differs from CCDs as, instead of an output register through which pixel data passes serially, they have an output amplifier for every column so that data from individual pixels can be read out in parallel. This has the advantage of significant speed increases over CCDs, but typically comes with the tradeoff of having a rolling shutter instead of a global shutter. Although global shuttering has the desired characteristic of instantaneous capture of an entire frame, a rolling shutter is only a liability when objects are moving across the field of view. Compensations for timing offsets between rows can easily be made during processing if deemed necessary. With a 25 ms exposure time and 2048 (2x2 binning) rows being read out, the row-to-row timing offset is 1.2 $\mu$s. Similar to the accuracy of our time stamp, it is not significant as long as the pointing differences between the three telescopes is small. Considering all of this, sCMOS-based cameras were the only technology that could meet the speed, field-of-view, and cost requirements of the Colibri project.

Although sCMOS-based cameras were the preferred choice, no suitable cameras were commercially available at the time of project conception. The cameras that were on the market were either equipped with small chips or delivered frame rates much lower than the required 40 fps. At the time, FLI was developing their Kepler series cameras which could be equipped with a large-format sCMOS sensor and, most importantly, would be able to image at frame rates \textgreater30 fps over a fiber connection. Because of this, FLI was chosen to supply three KL4040 cameras for the project.

The cameras have a number of possible modes of operation. They can be run in image capture mode where data is buffered and written to disk in several different formats, or they can stream straight from camera to disk. When running in the first mode, the cameras can be connected to the PC via either a USB3.0 or a fiber connection. When running over USB, speed is limited and we find a maximum frame rate of about 15 fps when 2x2 binned. A fiber connection allows for much higher frame rates (40+ fps), however the raw-to-fits format conversion when writing to disk slows things considerably after the camera's buffer has been filled. To allow for the required continuous high-speed imaging, data from the cameras is written directly to a RAID from a raw data stream with no format conversion. The raw format of the stream is one where each image container holds a header and both the low- and high-gain exposure data. The header takes up the first 246 bytes and contains necessary information about the exposure such as integration time and GPS time. After the header, the 12-bit high- and low-gain image data is interleaved by rows, giving a resultant raw image size of 12.6 MB. The two images can then be split and combined to create a single high dynamic range (HDR) 16-bit image. As the creation of an HDR image adds significant processing time, we currently extract only the high-gain image for processing. Because the use of the high-gain image gives roughly 2 magnitudes greater depth compared to the low-gain image without saturating stars in our target fields, trading off dynamic range and a slight increase in noise for processing speed has only a minor effect on our ability to detect SSOs with our pipeline.

\subsection{Network and Control Computers}
Two fiber networks connect the imaging equipment in the domes with control computers in a house located 270 m (along the fiber) from the furthest dome. The first network uses 1000BASE-SX transceivers over single-mode fiber and is used for communications with the telescope, focuser, guide and security cameras, and dome controller. The second network is used exclusively by the imaging camera. To keep up with the camera's high frame rate, variable bit-rate 40GBASE-CSR QSFP+ transceivers run at speeds of 8 Gb per second on each of four channels over 8 strands of OM3 multi-mode fiber. Speed tests have shown that frame rates up to 50 fps with 2x2 binned images are achievable, but are at the upper limit of our four-disk RAIDs.

Each dome has its own dedicated control computer to manage equipment operations, scheduling, and imaging tasks. The PCs are running Windows 10 with observatory control managed by ACP software. ACP allows us to connect to any device with ASCOM drivers which makes integration of off-the-shelf equipment relatively straight-forward. Technically, the PCs are mid-range machines with 3.2 GHz i7-8700 processors and 32 GB of RAM. To allow for image streaming to disk at rates of up to 4 Gb per second, each computer has a striped RAID of 4 10 TB Seagate Exos hard disks. This gives a peak performance of about 5 Gb per second with a total storage capacity of 36 TB. At 40 fps this gives enough space for about 20 hours -- roughly 2 to 3 nights -- of data.

\section{Colibri Data Handling}
\subsection{Data Acquisition}

Since the streaming mode of camera operation is not available through the FLI ASCOM driver, we have written our own camera control code in C\#. This lightweight application runs from the command line and gives the user access to the main camera settings while running in streaming mode. Command line usability also means that the program is accessible from within ACP's scripting environment.

The lack of ASCOM camera support for our operations also means that we need an alternative to ACP's scheduler. We accomplished this by writing JavaScript code to handle rudimentary scheduling that could be run from ACP. After initiation, the script checks for the presence of weather data. If present, the script continues. Otherwise, the user is given the opportunity to either quit or continue without the safety of the weather station data. The next step creates the data directories for that night's data and collects a set of 50 bias images. The script then calculates sunset and sunrise times, as well as the position of the moon relative to a  pre-determined list of stellar fields of observation. Details on the assembly of this pre-determined set of stellar fields are deferred to a future publication (Metchev et al.\ 2022, in preparation).

We iterate through each field in the list and estimate the number of visible stars according to airmass in 6 minute steps from sunset to sunrise. The fields are then filtered based on elevation (must be \textgreater10\textdegree) and moon angle (must be \textgreater15\textdegree), with ones that do not satisfy these criteria removed from the list of fields for the night. The remaining fields are then ranked, according to the estimated number of stars visible, for each 6 minute window. Once this has been done at the beginning of a night, a sunset check is made. If the sun is less than 12\textdegree\ below the horizon, the system pauses until the sun has passed this limit. 

Once it has been determined that it is dark enough to open, a weather-check is made and, if it is safe, the dome is opened, homed, and synced to the telescope. After another set of biases are collected, the telescope then slews to the first field and begins capturing data in subsets of 2400 images. Once another field surpasses the current field, the telescope is slewed to the new location and imaging begins once again. This is repeated until either sunrise or a weather alert shuts the system down. 

\subsection{Preliminary Data Processing}

After being written to disk, the images can be viewed individually with either FLI's Pilot software or with custom python scripts for viewing and conversion to FITS format. Batch conversion to FITS format averages about 20 ms per image. This can be done after a night of observations, but effectively takes nearly as long as the observations themselves. It is therefore an important limitation as we aim to process data during daylight hours. Our data storage allows up to a few consecutive clear nights of observations, which is generally sufficient given the $\sim$30\% fraction of clear nights at Elginfield Observatory.

As with CCD imagers, sCMOS cameras should be corrected for bias, flat-field, and dark current prior to quantitative/absolute work being done. 
Although the noise specifications are quite low, the sensors in our cameras have a considerable amount of bias structure that is especially visible during rapid imaging. As such, it must be removed by subtraction of a median combined set of bias frames.

With the short (25 ms) exposures that are typical for Colibri operation,  the low dark current (0.15 $\mathrm{e^-/pixel/s}$) of the sensor means that we can ignore separate dark current removal. Even then, because of the way that we collect the biases (25 ms, shutter closed), dark current is effectively removed during bias subtraction. Collection of a traditional bias with a 0 ms exposure is not possible currently due to limitations with the camera hardware.

The fast, wide-field nature of our optical system lends itself to vignetting on the image plane. This is visible as darkening outwards from the centre of the field and is easily corrected with a proper flat-field image. For stellar extraction and the relative photometry that we perform in our workflow, however, flat-fielding is not strictly necessary and is not applied during the first stage of processing due to time constraints.

\subsection{Occultation Detection Pipeline}
\label{sec:pipeline}

\begin{figure}[h!]
    \centering
    \includegraphics[width=\textwidth]{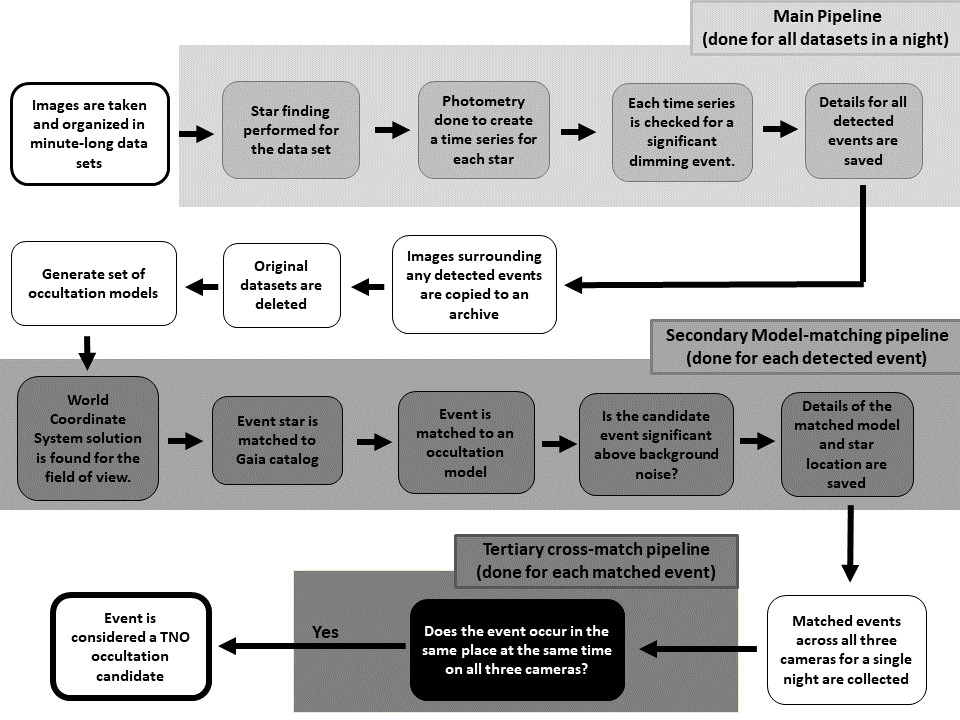}
    \caption{Processing pipeline flowchart showing the steps in the Colibri data reduction pipeline.}
    \label{fig:PipelineFlowchart}
\end{figure}

Although the detection pipeline (Figure \ref{fig:PipelineFlowchart}) follows the general description given by \citet{Pass2018}, changes to the equipment specifications have necessitated modifications to the pipeline. 

After the initial preparation of the data has been completed, stellar sources are extracted by running the Source Extraction and Photometry (SEP) module for Python \citep{Bertin-SExtractor1996, Barbary2016} on a median combination of the first nine images from each 2400 image subset collected over the night. The output from the SEP $extract$ function gives the coordinates for each star with a flux at least 4 $\sigma$ above the background at the start of each subset. After this, the average stellar drift within each subset is calculated. Every image in a subset is then interrogated using the $sum\_circle$ function of SEP to produce a time series of fluxes at each drift-corrected x,y location. The light curves are then passed to the first-stage dip detection where stars are filtered based on their SNR (must be greater than 5) and whether they drift out of the field. We then look for geometric dips greater than 40\% of the normalized light curve. A dip much less than 40\% is considered a common event (due to noise) and would represent a false detection. The geometric dip detections are automatically saved as candidates. In the case of no geometric dip,  the light curve is passed to the KBO dip detection function which convolves a Ricker wavelet designed to match the characteristic width of a KBO occultation event with each time series. The size of the wavelet is chosen based on the expected occultation duration given the pointing of the telescope. This means that, with expected durations of between 160 ms and 500 ms, Ricker wavelets of between 4 and 12 frames length are most appropriate. We then look for the most significant dimming events in each time series. If the dips of these events are greater than 3.75$\sigma$ of the mean of the convolved time series, the event is passed to the next stage of processing where the events are compared to a set of pre-calculated kernels that model diffraction patterns for a range of different physical/dynamical properties. Finally, successful matches from this step are checked for correlation in time with the other two telescopes of the array and saved offsite.

\section{System Performance}
\subsection{Optical}
\begin{figure}[h!]
    \centering
    \includegraphics[width=\textwidth]{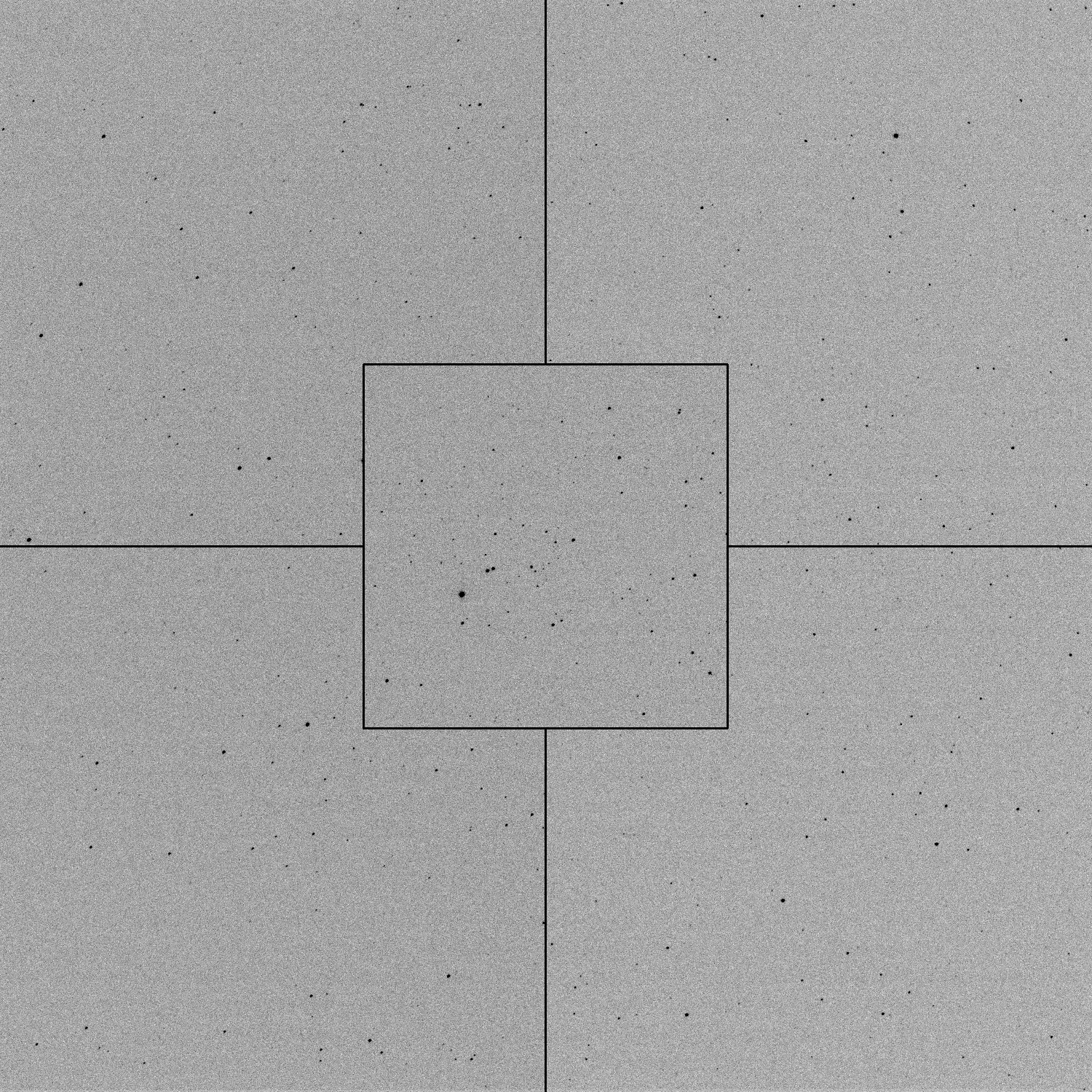}
    \caption{The four corners of this 10 s, $2\times2$ binned image (2.51$''$ per binned pixel) show spot sizes visually similar to the central portion of the image. The outer fields are 30$'$ square while the central square is 20$'$ square.}
    \label{fig:4CornerSpotSize}
\end{figure}

When pairing a fast (f/3) mirror design with an off-the-shelf (ASA Wynne) field corrector, the final results can be difficult to predict. The prescription for the ASA corrector is unknown, but the manufacturer claims that it has been designed with parabolic primaries from f/3--f/5 in mind. Extrapolating from their sales literature, we would expect spots ranging from about 4--12 $\mu$m (400--700 nm) on the optical axis to 12+ $\mu$m in the corners of our sensor. Although much better than an uncorrected system, our 9 $\mu$m pixels mean that we should be able to observe the difference in spot sizes from centre to sensor corners (Figure \ref{fig:4CornerSpotSize}).  Raytracing the as-built mirror specs with a generic Wynne corrector prescription, however, suggests that the design has excellent optical performance over the entire field for our observing conditions and our 2.5$''$ ($2\times2$ binned) pixels. With seeing often more than several arcseconds at the Elginfield site, we can tolerate slightly defocused images while still maintaining critically sampled PSFs across our $2\times2$ binned images.

Although the coma-corrected field-of-view is claimed to be 60 mm, there is vignetting at the field edges. This is visible in flat field imagery and amounts to a loss of 1 magnitude in the corners of the images. As a result, a flat-field correction should be done if we want to perform absolute photometry. Because of Colibri's limited processing power and the fact that we are doing same-star differential photometry, however, we do not flat-field correct our images within our processing pipeline.

\subsection{Mechanical}
After coarse polar alignment, the mounts were polar-aligned more precisely using the drift-alignment technique. Pointing models to 10 degrees altitude were then created using MaxPoint software from Diffraction Limited to quantify mount and telescope alignment. Mount alignment was then adjusted and new pointing models created in an iterative fashion until azimuth and altitude misalignments were below 10$''$.  After alignment and pointing model corrections, we find that the pointing performance of all telescopes is typically within a few 10s of arcseconds of the desired position. With our large fields, this is sufficient for ensuring that all three telescopes are monitoring the same stars.

\subsection{Photometry}
\begin{figure}[h!]
    \centering
    \includegraphics[width=\textwidth]{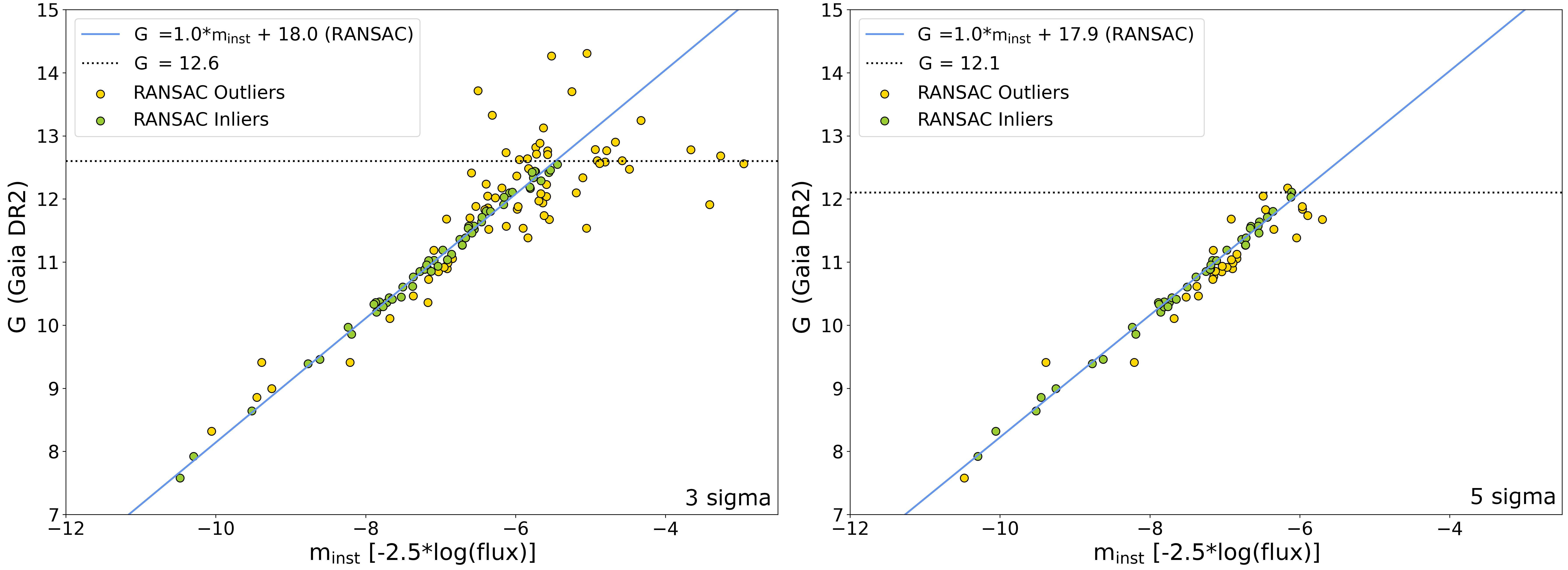}
    \caption{Gaia $G$-band vs.\ instrumental magnitude of 129 sources extracted with a detection threshold $\geq$3 sigma (left panel) and 69 sources extracted with a detection threshold of $\geq$5 sigma (right panel)} from a 25 ms exposure.
    \label{fig:MgVsMinst3and5sigma}
\end{figure}

When imaging rapidly with the KL4040 camera, background values can be seen to fluctuate by 10s of counts. To try to understand the effect that this has on rapid-imaging performance, we image a star-rich region of the sky (R.A.: 4.75h, Dec.: 72.75\textdegree) at our target frame rate of 40 fps without a filter. The images are corrected for bias and then stars are extracted using SExtractor \citep{Bertin-SExtractor1996}. A world coordinate system solution is obtained by processing an initial image of the chosen stellar field through the astrometry.net algorithm \citep{Lang2010}. Stellar positions are cross-correlated with their Gaia Early Data Release 3 \citep[EDR3;][]{Gaia2016b, Gaia2021} coordinates. Instrumental magnitudes are then compared with Gaia $G$-band (400--860~nm) magnitudes to derive a first-order transformation equation (Figure \ref{fig:MgVsMinst3and5sigma}).
%(Figures \ref{fig:MgVsMinst3sigma} and \ref{fig:MgVsMinst5sigma}). 
Using a Random Sample Consensus (RANSAC) algorithm to estimate the best fit to our data we find very similar solutions for each of the sigmas tested. One of the main benefits of the RANSAC solution is that misidentified stars or those with bad instrumental magnitudes will be classified as outliers and, as a result, have no impact on the final solution.

Stellar limiting magnitude has been defined \citet{Harris1990} as the magnitude at which only 50\% of the objects of that magnitude are detected. By injecting stellar sources with a range of magnitudes into real imagery, Harris counts the number of found sources with the  IRAF/{\sl DAOphot} software. The method we describe is slightly different, yet is still effective for our purposes. Instead of injecting synthetic stars into our images, we simply compare our extracted stars to known sources from the Gaia EDR3 catalogue 
%\citep{Gaia2016b, Gaia2021} 
to see at what limit we stop detecting objects. Figure \ref{fig:MgVsMinst3and5sigma} shows the instrumental magnitude of detected objects to a limit of 3 sigma above the background plotted against the Gaia $G$-band magnitude of the star closest to the extracted position. Beyond about $G=12.5$ mag, correlation between detections and real stars is lost and we set that as our practical limiting magnitude for a 25 ms exposure.

Although an SNR of 3 is considered to be the absolute limit for most systems, the actual limit for observing occultation events is somewhat higher.  For the detection of KBOs, we set our stellar detection threshold to 4 sigma but realistically expect to be able to detect SSOs down to about 5 sigma above the background. We use the fitted relation in Figure~\ref{fig:MgVsMinst3and5sigma} to calibrate our conversion from Colibri instrumental to Gaia $G$-band magnitudes. Since Colibri's is a filter-less system, we do not seek to accurately transform the data to a standard photometric system. Our conversion is likely accurate to $\pm$0.1~mag for most stars.

\begin{figure}[h!]
    \centering
    \includegraphics[width=\textwidth]{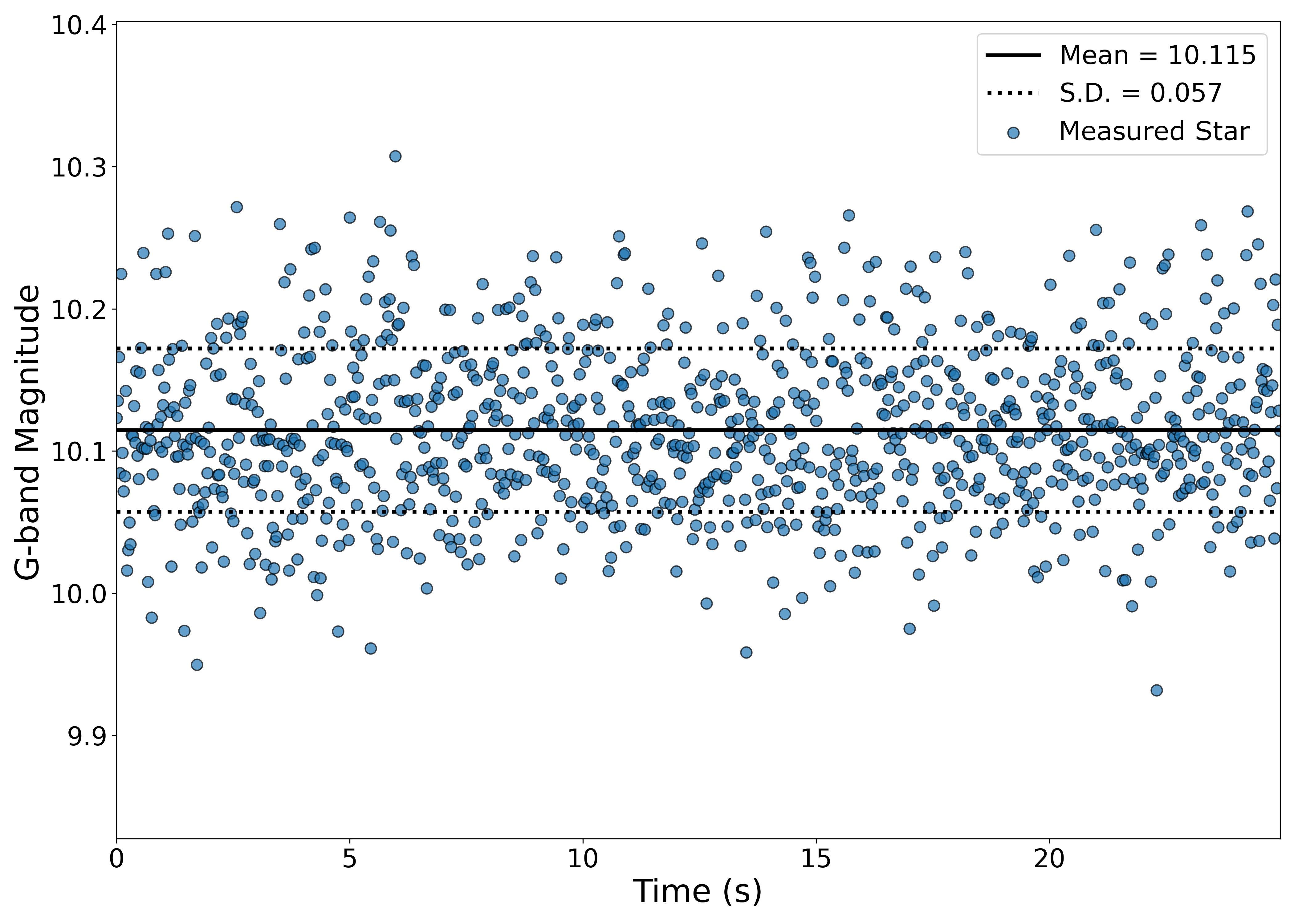}
    \caption{A magnitude vs.\ time light curve for a $G=10.1$ mag star over a 25 second window in one thousand 25 ms exposures.}
    \label{fig:MagVsTime}
\end{figure}

Using the same 40 fps dataset as above, we also examine the relative photometric temporal stability of the system. To do this, we look at the frame-to-frame residual changes in measured magnitude (instrumental or transformed) for all stars in the field. Figure~\ref{fig:MagVsTime} shows the measurements for a randomly-selected $G=10.1$ mag star over 1000 consecutive 25 ms exposures (25 sec sequence). The standard deviation is 0.057~mag. We use a derivative measure, the ratio of the mean flux to the r.m.s.\ scatter of the flux over a 60-sec interval, as the ``temporal SNR'' of our flux measurements. The temporal SNR is approximately diagnostic of the minimum detectable depth for a stellar occultation. For example, at a temporal SNR of 10 the rms scatter of the light curve is $1/10 = 1\%$, so detectable occultation would be at least $\sim$30--40\% deep. Our first dip-detection pass in our occultation detection pipeline does indeed set a 40\% threshold for detection (Section~\ref{sec:pipeline}).

\begin{figure}[h!]
    \centering
    \includegraphics[width=\textwidth]{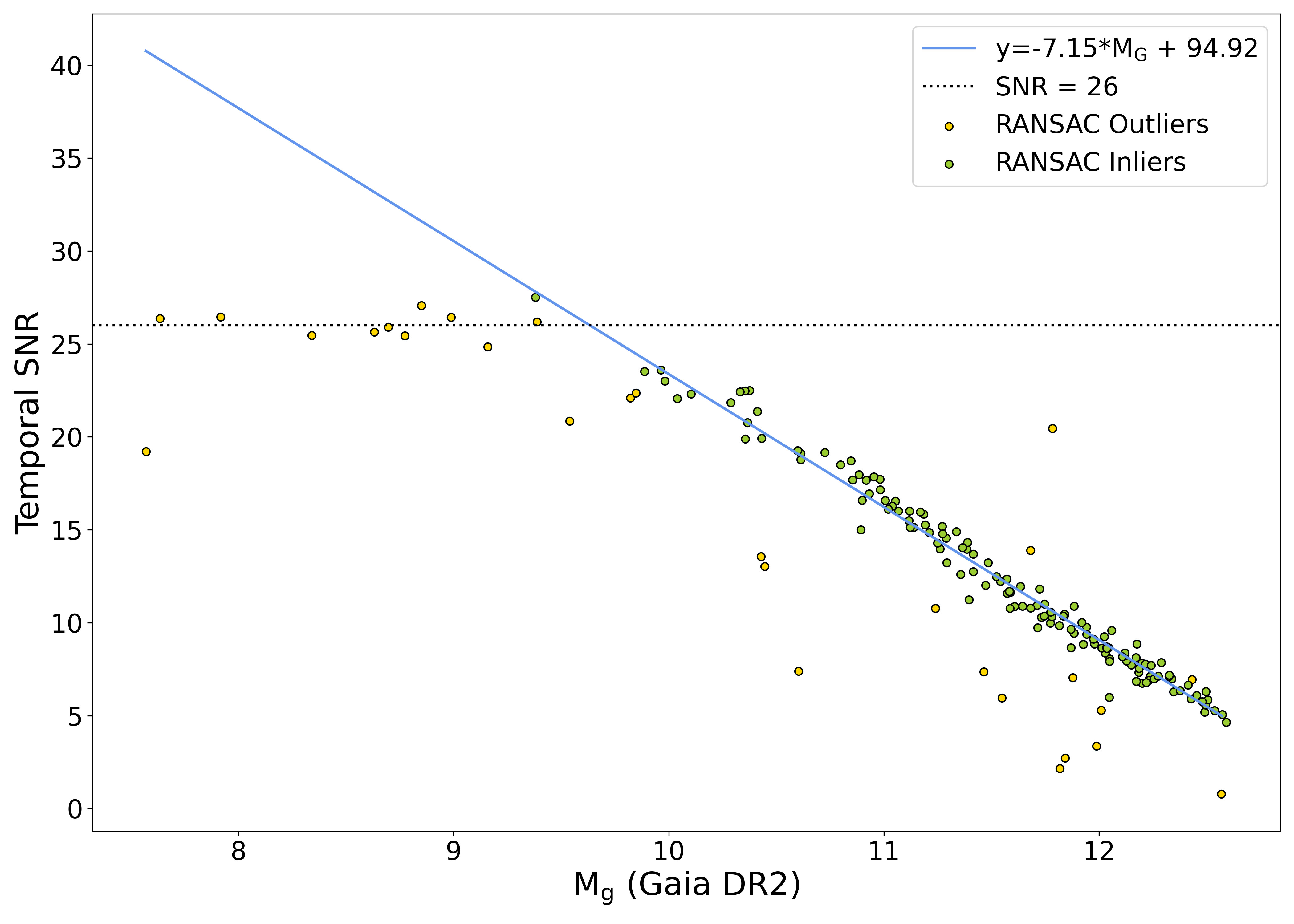}
    \caption{The temporal SNR of $G\lesssim10$ mag objects, extracted from a field centred at an airmass of 1.37, is roughly SNR=26. At fainter magnitudes, the temporal SNR decreases linearly to SNR = 5 at $G\approx12.5$ mag.}
    \label{fig:SNRvsMg}
\end{figure}

Figure \ref{fig:SNRvsMg} presents the temporal SNR of objects (airmass $\approx$ 1.37) with 25 ms exposures as a function of stellar $G$ magnitude. The temporal SNR follows a linear trend for stellar magnitudes fainter than about $G=9.5$~mag. Brighter objects, however, are best described by a horizontal line at an SNR of about 26. This is likely an indicator that atmospheric scintillation becomes the dominant source of noise for bright objects.

When the standard deviations of the measurements for all stars against their mean magnitudes are plotted (Figure \ref{fig:SNRvsMg}), they show a monotonically increasing standard deviation with increasing stellar magnitude. This provides a useful estimate of the threshold for the depth of a detectable dip at a given magnitude. The 0.035 mag error ($G\lesssim9.5$~mag stars), for example, corresponds to 3.5\% of the flux while a 0.2 mag error ($G=11.5$ mag stars corresponds to 17\%) of the total flux. Therefore, we can conclude that most \textgreater20\%-deep occultations of brighter stars should be visible. Occultations of fainter stars are more likely to be detected under ideal conditions or if $\gtrsim$50\% depths occur under favourable occultation geometry. As already detailed in \citet{Pass2018}, stellar brightness will be critical for the precision to which the parameters of a stellar occultation can be determined.

\subsection{Astrometry}
\begin{figure}[h!]
    \centering
    \includegraphics[width=\textwidth]{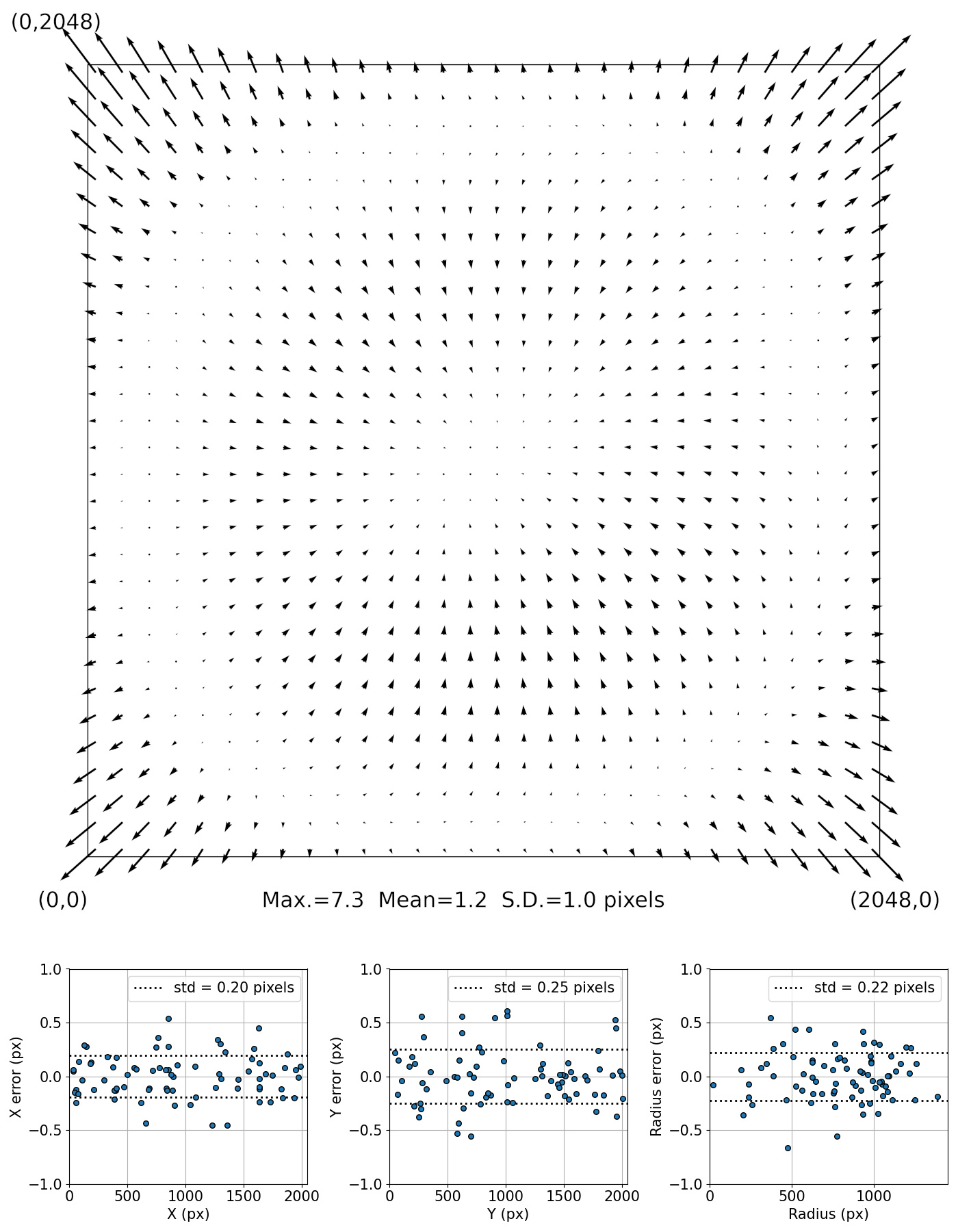}
    \caption{Distortion (scaled 20x) as computed using a third-order polynomial with a first-order radial term is shown in the upper panel. The lower panel shows the astrometric residuals after correction. }
    \label{fig:AstrometryFit}
\end{figure}

 Using the SkyFit2 tool \citep{Vida2021-RMS}, a plate solution is computed using a third-order polynomial with a first-order radial term (Figure \ref{fig:AstrometryFit}). When compared to Gaia EDR3, the results show the standard deviation of the residuals ranging from about 0.2 to 0.25 pixels (0.5$''$ to 0.63$''$) with no obvious dependence on distance from the field centre. Higher order fits have been tested, but there is no obvious benefit to using them given the low amount of distortion (Figure \ref{fig:AstrometryFit}, upper panel) in the system and the fact that high astrometric precision is not a strict requirement for the detection of KBOs by stellar occultation. Although this represents a solution from a single pointing direction (altitude=45\textdegree, azimuth=0\textdegree), we don't find a large changes in the plate solution at different pointings.

\section{Summary}
The Colibri telescope array at Elginfield Observatory in Ontario, Canada, has been built for the purpose of detecting kilometre-sized KBOs by the serendipitous stellar occultation method. The construction used cost-effective, off-the-shelf components to meet the design goals of the Colibri project. 

Using sCMOS cameras, Colibri has been designed to continuously monitor the night sky for serendipitous stellar occultation events. At 40 fps, each camera streams imagery at 4 Gb/s over a 40G fibre link to its own dedicated RAID. This generates up 20 TB of data per camera per night to be processed the following day by the processing pipeline.

The photometric and astrometric performance of the system has been measured. Photometrically, the system performs well with a limiting broad-band $G$ magnitude of about 12.5 (temporal SNR=5) in a 25 ms exposure. The relationship between instrumental and Gaia $G$ magnitudes is linear with a 1:1 slope while light curve analysis shows stability in the measurements within the timescale of the 1-minute data subsets. The standard deviation in the magnitude measurements increases with increasing magnitude, ranging from about 0.035 mag on $G\lesssim9.5$~mag stars to $\sim$0.2 mag at the SNR=5 limit. As this corresponds to a roughly 3.5\% to 17\% variation in flux, occultations with $\sim$50\% flux depths over several consecutive exposures should be detectable at even the faintest magnitudes. Astrometrically, the optics show some distortion and better than $\pm$0.25 pixel ($\pm$0.6$''$) errors. As a result, a third-order polynomial plate solution is sufficient for this project. 

The Colibri observatory is currently collecting a limited amount of data while testing the automation and processing pipeline routines. Full operational activities are expected to begin in the summer of 2022.

\bibliographystyle{Frontiers-Harvard}
\bibliography{ColibriFrontiers}

\end{document}